\documentclass[journal]{IEEEtran}
\IEEEoverridecommandlockouts
\usepackage{amsmath,amssymb,amsfonts}

\usepackage{graphicx}
\usepackage{textcomp}
\usepackage{xcolor}
\usepackage{comment}

\usepackage{array}
\usepackage{dcolumn}    
\usepackage{bm}         
\usepackage[normalem]{ulem}
\usepackage{amsmath}
\usepackage{amsthm}
\newtheorem{theorem}{Theorem}
\usepackage{algorithmicx}
\usepackage{algorithm}
\usepackage{algpseudocode}
\usepackage{gensymb}
\usepackage{siunitx}
\makeatletter
\def\BState{\State\hskip-\ALG@thistlm}
\makeatother
\usepackage{soul}
\usepackage[nolist,nohyperlinks]{acronym}
\usepackage{upgreek}
\usepackage{subfigure}
\usepackage{epsfig}
\usepackage{caption}
\usepackage{tikz}

\def\BibTeX{{\rm B\kern-.05em{\sc i\kern-.025em b}\kern-.08emT\kern-.1667em\lower.7ex\hbox{E}\kern-.125emX}}
    
\begin{document}

%

\title{Toward Fault-Tolerant Deadlock-Free Routing in HyperSurface-Embedded Controller Networks}

\author  {Taqwa Saeed, \IEEEmembership {Member IEEE}, Vassos Soteriou, \IEEEmembership {Senior Member IEEE}, Christos Liaskos, \IEEEmembership {Member IEEE}, Andreas Pitsillides, \IEEEmembership {Senior Member IEEE}, Marios Lestas, \IEEEmembership {Member IEEE}

\thanks{Taqwa Saeed and Andreas Pitsillides are with the Computer Science
Department, University of Cyprus, Nicosia, Cyprus {\tt (e-mail: tkhair01@cs.ucy.ac.cy}, {\tt andreas.pitsillides@ucy.ac.cy).}} 

\thanks{Vassos Soteriou is with the Department of Electrical Engineering and Computer Engineering and Informatics, Cyprus University of Technology, Limassol, Cyprus{\tt(e-mail: vassos.soteriou@cut.ac.cy)}}
\thanks{Christos Liaskos is with the Foundation of Research and Technology (FORTH), Heraklion, Greece {\tt(e-mail: cliaskos@ics.forth.gr).}}%

\thanks{Marios Lestas is with the Electrical Engineering
Department, Frederick University, Nicosia, Cyprus {\tt(e-mail: eng.lm@frederick.ac.cy).}}
\vspace{-5 mm}

}

\maketitle

\begin{abstract}
HyperSurfaces (HSFs) consist of structurally reconfigurable metasurfaces whose electromagnetic properties can be changed via a software interface, using an embedded miniaturized network of controllers. With the HSF controllers, interconnected in an irregular, near-Manhattan geometry, we propose a robust, deterministic Fault-Tolerant (FT), deadlock- and livelock-free routing protocol where faults are contained in a set of disjointed rectangular regions called faulty blocks. The proposed FT protocol can support an unbounded number of faulty nodes as long as nodes outside the faulty blocks are connected. Simulation results show the efficacy of the proposed FT protocol under various faulty node distribution scenarios.
\vspace{-3mm}

\end{abstract}
\section{Introduction}\label{sec:intro}
Large Intelligent Surfaces~\cite{nadeem2019large} have recently attracted momentous attention promising to revolutionize wireless communications, redefining even their fundamental principles \cite{basar2019reconfigurable}. Software-Defined Metasurfaces~\cite{Liaskos2015design} (SDMs) comprise a vital candidate enabler technology for the implementation of Large Intelligent Surfaces, possessing advanced properties relative to competing technologies~\cite{tan2018enabling}. 
In recent times, considerable effort has been spent in complementing the underlying metasurface hardware with a complete set of protocols and an Application Programming Interface (API), realizing \textit{HyperSurface} (HSF) units ready to be incorporated in applications, such as Programmable Wireless Environments (PWEs)~\cite{liaskos2019network}. 

\begin{figure}[t!]
	\centering
		\includegraphics[width= \columnwidth]{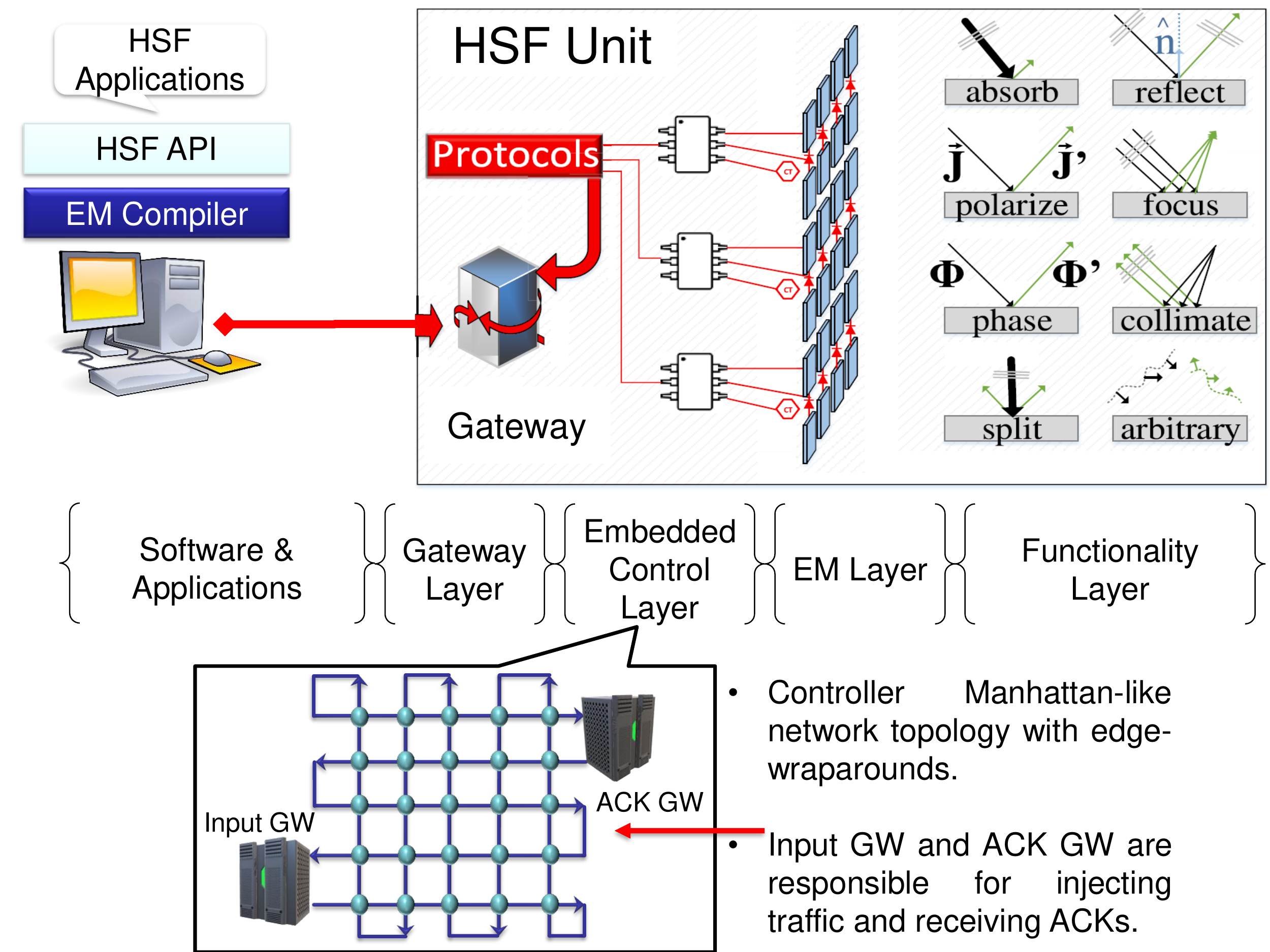}
	\caption{\small Layered HyperSurface (HSF) system.}
	\label{sys}
	\vspace{-6mm}
\end{figure}

HSF operation and structure are defined by a five-tier layered model \cite{Abadal.2017}, depicted in Fig.~\ref{sys}. First, the functionality layer comprises the types of electromagnetic (EM) control functions that can be exerted by the HSF over an impinging wave, while the EM layer consists of a periodically repeated element, the \textit{meta-atom},   comprising state-altering material of a specific geometry~\cite{pitilakisMETA2018}. Next comes the Embedded Control Layer, described later, while the gateway layer connects the HSF to the outer world relaying required external states to the internal embedded state-altering elements, as well as HSF monitoring information to the opposite direction. Last, the EM compiler acts as middleware, transforming API callbacks to actual actuation directives that configure the HSF controllers and the associated meta-atoms.
The HSF embedded control layer is stationed between the gateway and the EM layers~\cite{ISCAS18}, constituting the necessary hardware and protocols establishment to transfer data between the gateway and the state-altering elements (DC voltage-controlled, continuously tunable varactors/varistors). In its simplest form, the control layer comprises direct wiring of the state-altering elements to the gateway, sufficient even for some PWE application scenarios~\cite{liaskos2019network}. However, as the density of meta-atoms increases, a \textit{network of embedded data controllers} is required to exert control within the HSF for online re-configuration purposes. The design of this control layer is challenged by factors such as low power consumption, dense deployment, the small size of meta-atoms, EM interference avoidance, and
robustness against any type of fault, etc. These challenges \cite{kouzapas2019towards} have led to a Controller Network (CN) topology which is irregular and near-Manhattan, as shown as part of Fig.~\ref{sys} in blue color, and in Fig.~\ref{fig:TopologyRouting}, where the directionality of rows (and columns) alternates from one row (column) to the immediate next, resembling a mesh topology where approximately half of its unidirectional links are missing. Reconfiguration Directives (RDs) for each interconnected meta-atom are inserted via a \underline{G}ate\underline{W}ay (GW) connected to the lower left periphery of the network, while once a RD datum is received by a meta-atom, then its associated controller sends an \underline{ACK}nowledgement (ACK) message to the ACK GW connected at the topmost right corner of the CN.

In terms of miniaturization and layout, there exist resemblances between Network-on-Chip-interconnected (NoC) multicore processor architectures and the geometry of HSF-embedded CN. However, the HSF CN \textit{suffers from severe resource limitations}, and hence exhibits a plethora of distinct operational characteristics vs. most conventional NoCs, including a unique unidirectional interconnect geometry and clock-less operation~\cite{petrou2018asynchronous}, though some NoC routers do function asynchronously~\cite{asynchronousNocRostislav}, detailed in~\cite{kouzapas2019towards}. As such, the HSF CN methodologies that we develop here to mitigate targeted challenges faced by the HSF CN, cannot be directly inherited from NoC designs; instead, customized solutions become necessary.

HSFs containing an embedded CN are complex devices with densely integrated tuning elements, unit cells, controller chips, etc., in non-standard packages. Such HSFs must safeguard against faults during fabrication, deployment, or operation, which have been shown to degrade HSF performance and are caused by factors such as connector misalignment, ageing, physical or intentional damage of unit cells, etc.~\cite{taghvaee2020error}. As such, to establish dependable communication among interconnected controllers in an HSF structure, we propose a link-level Fault-Tolerant (FT) routing algorithm that arms routed packetized directives\footnote{We use the terms ``software directives,'' ``directives,'' and ``data'' interchangeably throughput this paper.}, that reconfigure EM properties of meta-atoms in the HSF, in bypassing faulty nodes on their way to their destination controllers. Radetzki’s et al. survey~\cite{radetzki2013methods} provides a broad coverage in the field of FT approaches in NoCs, and the interested reader is directed therein. Essentially, as long as FT routing provides full connectivity devoid of cyclic channel dependencies in a sub-connected topology, then the FT function crucially guarantees packet delivery in deadlock-free\footnote{We utilize the term ``deadlock-freedom'' as used in the seminal work by  Duato~\cite{DuatoDeadFreeTheory} all over this paper; that is, message traversals along the CN network are devoid of cyclic patterns as no circular channel (i.e., link) dependencies are allowed. As a result, any messages cannot be involved in a deadlocked situation, occurring, either (1) between packetized software directives, or (2) between such directives and acknowledgement packets (ACKs), or (3) among ACK packets, all scenarios which may halt the flow of messages indefinitely and render the CN of the HSF inoperable (refer to Section~\ref{sec:general_R}).} and livelock-free mode~\cite{DuatoDeadFreeTheory}. Designing a routing protocol that is both FT and livelock-/deadlock-free raises major challenges, especially in the case of the HSF network where the controllers are interconnected with said near-Manhattan topology, which possesses about $50\%$ fewer links vs. 2-D mesh-connected NoC topologies, in tandem with minimal HSF CN resources. In this work, we focus on devising a deadlock-free routing protocol that is \textit{strongly} coupled to said HSF CN topology so as to adhere to the above deadlock-/livelock-free routing design requirements, tailored to the needs of the specific environment. 

The adopted design methodology, assumes the formation of disjointed rectangular regions called Faulty Blocks (FBs), initially proposed by Wu~\cite{wu2003fault}, and utilizes principles from the Turn Model for adaptive routing by Glass and Ni~\cite{glass1992turn}. Based on assumptions outlined in Section~\ref{sec:general_R}, the resultant FT routing protocol we build for the HSF CN meets the following ideal objectives~\cite{wu2003fault}: 1) it is \textit{distributed}, with all routing decisions made locally at nodes to conserve resources, 2) it is \textit{feasible}, ensuring reliable delivery of directives to nodes and ACK messages to the ACK GW, 3) it is \textit{fault-tolerant} as it is able to bypass faulty nodes contained in FBs, 4) it is \textit{devoid of livelocks and deadlocks}, 5) it comprises a \textit{reasonable} fault model using FBs, 6) it ensures that routed directives span \textit{short routes} when bypassing nodes within FBs, and 7) it is \textit{adaptive} as it conforms to FB formations. Simulation results demonstrate the effectiveness of the proposed FT routing protocol under two fault distribution scenarios (see Section~\ref{sec:simulations}).



\vspace{-1mm}
\section{\label{sec:general_R}Fault-Tolerant Routing in the HSF CN}
\vspace{-0mm}
We progressively build our FT routing protocol suited for the HSF CN topology by concurrently considering 1) the flow of traffic dictated by link interconnectivity, 2) destination node positioning, and 3) the placement of the input and ACK GWs. 
\vspace{-3mm}
\subsection{XY-YX Agnostic Routing for a Fully-Healthy HSF CN}\label{subsec:XYAgonsticRouting}

\begin{figure}[t!]
	\centering
		\includegraphics[width= .6\columnwidth]{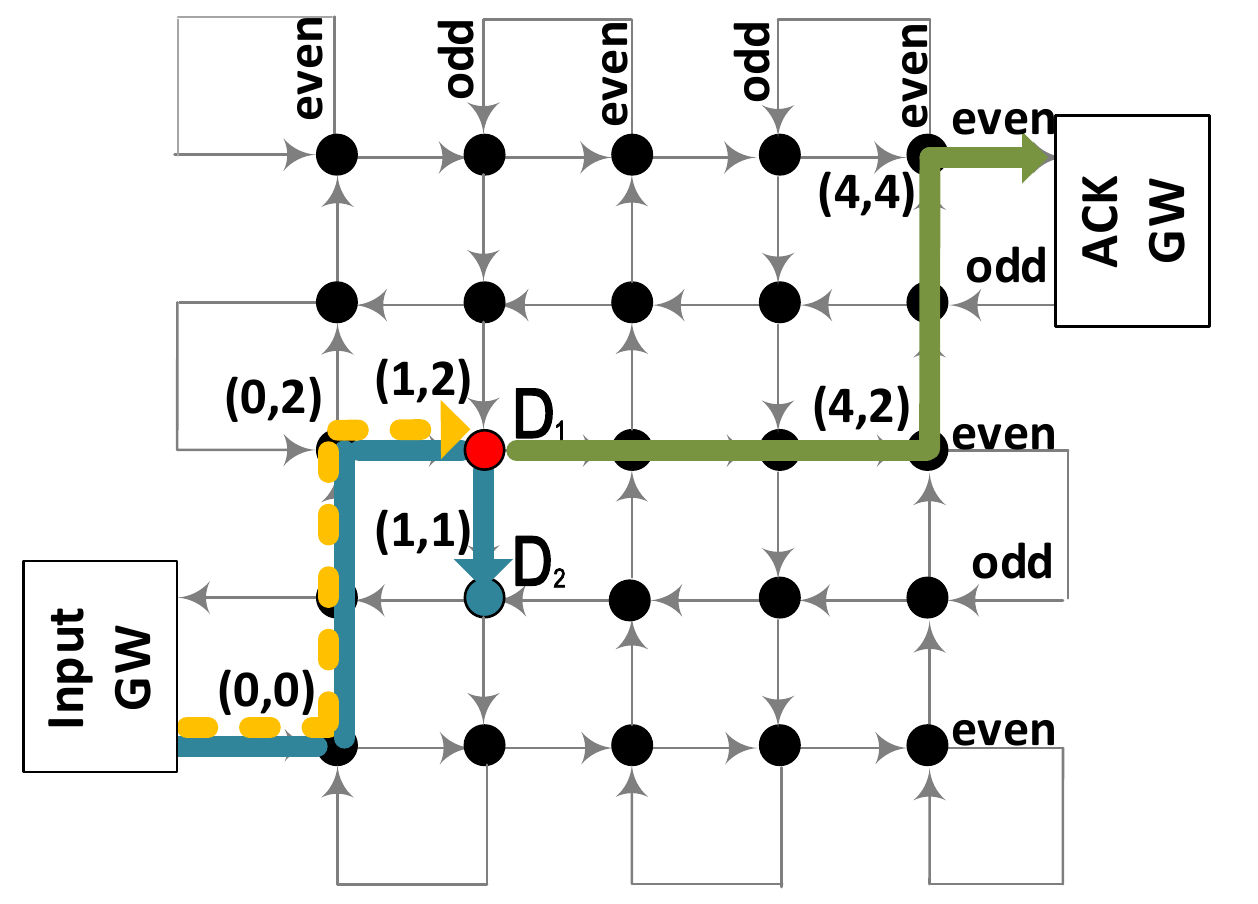}
		\vspace{-1mm}
	\caption{\small A $5 \times 5$ HSF-embedded controller network topology with two packetized directives routing scenarios shown in solid blue and dotted orange links; the ACK path is shown in solid green.}
	\label{fig:TopologyRouting}
	\vspace{-5mm}
\end{figure}

Regular XY-YX routing, built to suit a fully-connected HSF CN topology \cite{kouzapas2019towards}, must be properly adjusted to \textit{strongly adhere to the geometrical idiosyncrasy} of the HSF CN to reliably deliver directives in a deadlock-/livelock-free mode. As such, topology awareness is incorporated into base XY-YX routing in a \textit{distributed} fashion, with the method dubbed \textit{agnostic XY-YX} routing. To explain its workings, we make use of a demo shown in Fig.~\ref{fig:TopologyRouting}, where two directives are inserted from the input GW at coordinates ($x=0$, $y=0$), with one destined for node $D_{1}$ residing at ($1$,$2$), and the other destined for node $D_{2}$ at ($1$,$1$). The routing rules for directives are as follows: upon injection at the input GW, 1) if the destination lies on an even column, route horizontally until the $x$ coordinate offset reaches zero and then turn northwards, otherwise, 2) for an odd column destination, follow 1) but make a turn up a hop earlier ($x$ offset = $1$) and route northwards, 3) once step 1 or 2 is complete, when the $y$ offset becomes zero and the row is even, then make a turn to the east and route toward the destination node, otherwise, 4) hop one row further to the north, then make a turn and route eastwards until the $x$ offset is zero and then traverse one hop south to reach the destination node. In Fig.~\ref{fig:TopologyRouting} the directive taking the dotted orange route toward $D_{1}$ utilizes rules 2 and 3 (route: ($0$,$0$) $\rightarrow$ ($0$,$2$) $\rightarrow$ ($1$,$2$)), while the blue route toward $D_{2}$ utilizes rules 2 and 4 (route: ($0$,$0$) $\rightarrow$ ($0$,$2$) $\rightarrow$ (1,2) $\rightarrow$ ($1$,$1$)). The above topology-agnostic XY-YX algorithm is proven to be deadlock-/livelock-free as no full \ang{360} turns are formed which would produce either cyclic dependencies (deadlocks) or endless cycles in the topology (livelocks). The ACK packet generated by destination (1,2) follows a simple XY path toward the ACK GW; to avoid deadlocks, the ACK packet for destination node (1,1) is sent by its neighboring node at (1,2) (which is aware of the non-faulty status of node (1,1,) through their link interconnectivity) to avoid a forbidden cycle formation, and also follows the green path in Fig.~\ref{fig:TopologyRouting}.

\begin{figure}[t!]
	\centering
		\includegraphics[width=.46\textwidth]{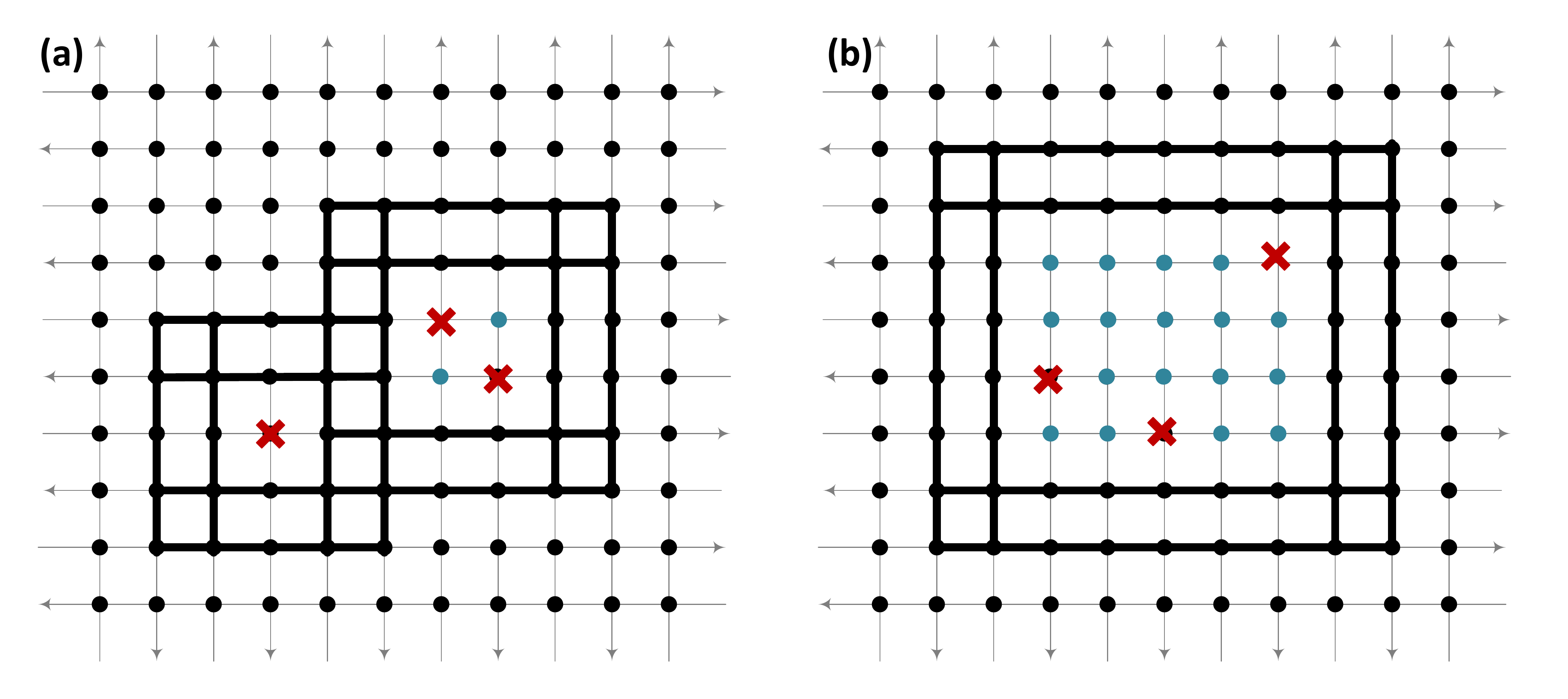}
		\vspace{-3mm}
	\caption{\small Faulty block formations demos with their boundaries marked in bold lines; faulty nodes are marked with an ``X'' sign, while unsafe and healthy nodes are respectively marked in blue and black.}
	\vspace{-5mm}
	\label{fig:faultyBlocks}
\end{figure}

\vspace{-2mm}
\subsection{Faulty Blocks (FBs) Formation Rules for FT Routing} \label{subsec:FB_formation}
\vspace{-1mm}



Prior-art in the 2-D mesh interconnection network domain has struggled to produce ample FT routing schemes that \textit{simultaneously} satisfy crucial requirements such as message delivery guarantee to connected nodes, deadlock- and livelock-freedom in routing, large fault counts handling, and reasonable implementation~\cite{DuatoBook, radetzki2013methods,DuatoDeadFreeTheory}. Unfortunately, also here we are faced with the same challenges: the elegant XY-YX agnostic routing algorithm cannot accommodate faulty nodes in the HSF CN, especially when randomly placed. Conducted simulations with such fault placements showed the XY-YX scheme breaking down as packets reached dead ends, with no turn options to avoid such obstacles in a distributed manner. 


\begin{figure}[t!]
	\centering
		\includegraphics[width=.14\textwidth]{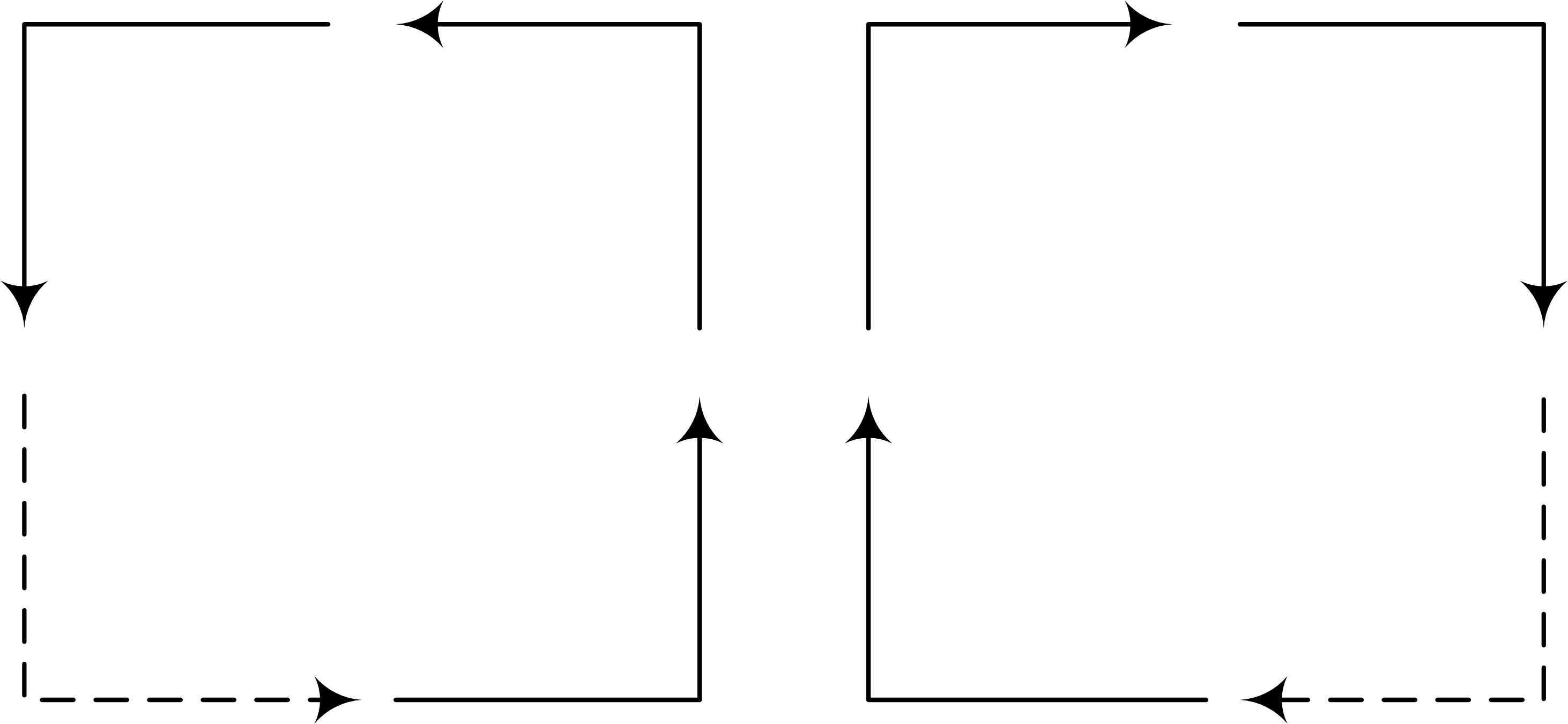}
		\vspace{-1mm}
	\caption{\small South-last routing with prohibited turns in dotted lines.}
	\vspace{-5mm}
	\label{fig:southlastturns}
\end{figure}


\begin{figure}[t!]
	\centering
		\includegraphics[width=.32\textwidth]{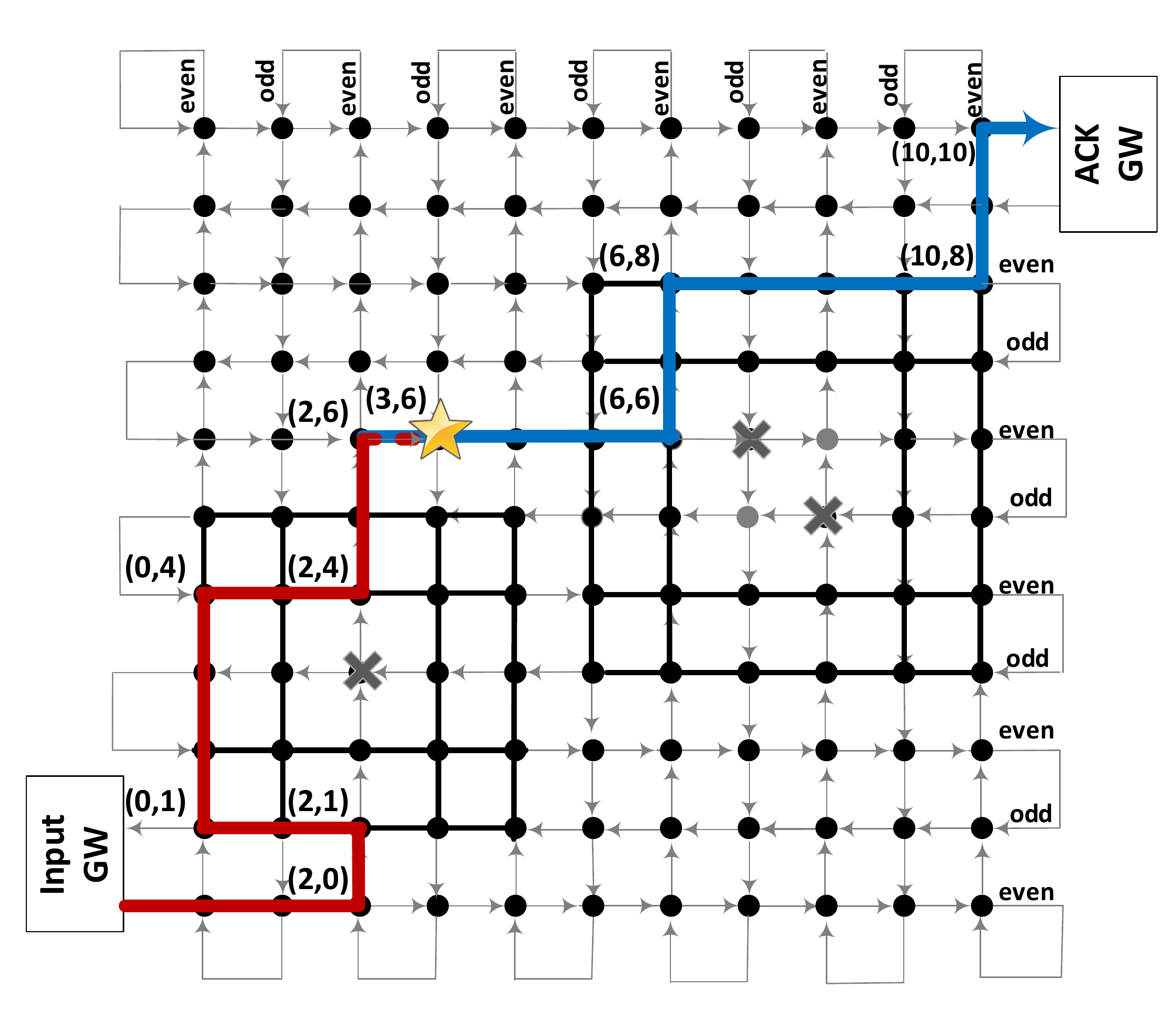}
		\vspace{-3mm}
	\caption{\small Routing around faulty blocks: directives (red), ACKs (blue).}
	\vspace{-6mm}
	\label{fig:rrouts}
\end{figure}

Fortunately, in our quest of achieving all above FT Routing Algorithm (RA) characteristics in the resource-limited HSF CN, we made an important observation: that packetized directives and ACKs which route toward the north-east CN corner, never use  the  edge  wraparound  links (they may be used in future HSF technology node implementations where controllers may also become producers~\cite{liaskos2019network}), and rarely travel westbound along odd rows. All these, along with the fact that routing and network topology are strongly coupled~\cite{DuatoDeadFreeTheory, DuatoBook}, prompted us toward the derivation of a FT RA suitable for the HSF CN where all faulty nodes are ``spatially bounded'' within simple geometrical shapes. Such setup would effectively support routes toward the north-east corner of the HSF CN topology while eliminating south-west turns to ensure deadlock-freedom (see Section~\ref{subsec:turnRules}). As such, we decided to utilize the idea of \textit{faulty blocks}, initially proposed for 2-D mesh interconnects~\cite{wu2003fault}, where faults are constrained in disjointed rectangular regions. Such FBs can contain \textit{``victimized''} non-faulty nodes that are marked as \textit{unsafe} and are hence unreachable, so as to maintain FB convexity and facilitate simpler FT routing; node classifiers also include \textit{safe} (i.e., healthy), \textit{faulty}, and \textit{boundary} nodes (that are healthy) spanning the periphery of FBs. We assume that the input GW identifies faulty nodes dynamically, marking the class of each such node. A node is declared unsafe according to these two rules: 1) it has two unsafe or faulty 1-hop neighbors, and 2) it has two unsafe or faulty 1-hop or 2-hop neighbors in two different dimensions.


As the two FB examples in Fig.~\ref{fig:faultyBlocks} show, a FB comprises four boundaries, each consisting of two adjacent rows or columns. The reason for having double boundary ``lines'' is to overcome the uni-directionality of the network links by including two ``lines'' of opposite directions for each boundary. The above two rules ensure that no two FBs can intersect without forming a larger FB. This can be proven by the following theorem. 

\vspace{-2mm}
\begin{theorem}
The boundary nodes of a faulty block cannot intersect with another faulty block.
\vspace{-1mm}
\end{theorem}

\vspace{-3mm}
\begin{proof}
Assume that node $v$ belongs to FB $A$. According to the definition of faulty blocks, $v$ must be an unsafe or a faulty node. Thus, it cannot be a boundary node of any other FB. This proves a contrapositive, which infers that the boundary node of a specific FB cannot belong to any other FB.
\end{proof}
\vspace{-2mm}

\vspace{-1mm}
\subsection{HSF FT Routing: Turns Employed \& Prohibited}\label{subsec:turnRules}
\vspace{-1mm}




To achieve the objectives of designing a FT, distributed, deadlock-/livelock-free, feasible, adaptive, short-route routing algorithm, the following assumptions are used in this paper: 1) only node faults are considered, with all four links at a controller considered faulty and inoperable, 2) only permanent dynamically-occurring faults detected by the input GW are considered (though we assume that no new faults occur during routing or FB marking), 3) that faulty nodes can neither reside on the periphery of the CN topology, nor on the row and columns adjacent to the network northern, western and eastern peripheries, while in addition, faults can neither reside on the row adjacent to the southern periphery nor on the row above the former, 4)  that the destination nodes always reside outside of faulty blocks, and 5) no limit to the number of faulty nodes exists as long as assumption 3) holds and that all faults are contained withing FBs according to FB formation rules.

As explained in Section~\ref{subsec:FB_formation}, due to the unidirectionality of the HSF CN links and the positioning of the two GWs, directives (and ACKs) collectively route toward the north-east (NE) direction, and rarely span westbound links. Hence going south comprises the \textit{last resort} routing direction. As such, and in tandem with the well-defined convex FBs proposed in Section~\ref{subsec:FB_formation}, and using the deadlock-avoidance blueprint dictated by the Turn Model of adaptive routing~\cite{glass1992turn}, where at least a turn in the clockwise and anti-clockwise routing directions is forbidden so as to eliminate deadlock-inducing \ang{360} turns in the absence of virtual channel usage, we prohibit south-west (SW) and south-east (SE) turns altogether, to form \textit{south-last} (SL) routing as shown in Fig.~\ref{fig:southlastturns}. SL is regarded as a \textit{subset} of the XY-YX RA (see Section~\ref{subsec:XYAgonsticRouting}) in the wider context, hence once a fault appears in the CN essentially our topology-agnostic RA becomes tighter. Specifically, we block SW turns on odd row and column intersections, and SE turns at odd columns intersecting with even rows; the only time a packet turns south for a last one-hop traversal is when the destination is on an odd column and an odd row; ACKs remain unaffected as they always also route in the NE direction.

The strong coupling of said FBs and the SL RA scheme produces a deterministic, distributed and more importantly a provably deadlock-/livelock-free FT routing mechanism. Deadlock-freedom can be proven either using theorems such as the ones formulated by Duato~\cite{DuatoDeadFreeTheory,DuatoBook} which are based on set theory, or by using brute-force simulations; both were carried out and omitted here for brevity. To showcase the workings of our HSF CN algorithm we make use of the example demonstrated in Fig.~\ref{fig:rrouts}. A packetized directive (red route) is inserted in the CN via the input GW at coordinates ($x=0$,$y=0$) heading toward destination node (yellow star) at ($3$,$6$). Due to the distributed nature of our FT routing scheme and based on the rules outlined in Section~\ref{subsec:XYAgonsticRouting} it makes an east-north (EN) turn at ($2$,$0$) where it encounters the boundary of the first FB at ($2$,$1$); it then misroutes two hops to the west (left) after a north-west turn and then routes along the periphery of the FB, eventually making an EN turn at (2,4) toward its destination. Node ($2$,$6$) then creates an ACK packet which is routed to the ACK GW (blue line), making an EN turn once the boundary of the second FB is met at ($6$,$6$). It then routes along the periphery of the same FB until it is successfully received by the ACK GW. No turn rules were violated, hence deadlock-freedom in routing is established.

\subsubsection{Deadlock Avoidance with Super-Faulty Blocks (SFBs)}\label{subsubsec:BlocksOvelap}

\begin{figure}[t!]
	\centering
		\includegraphics[width=.4\textwidth]{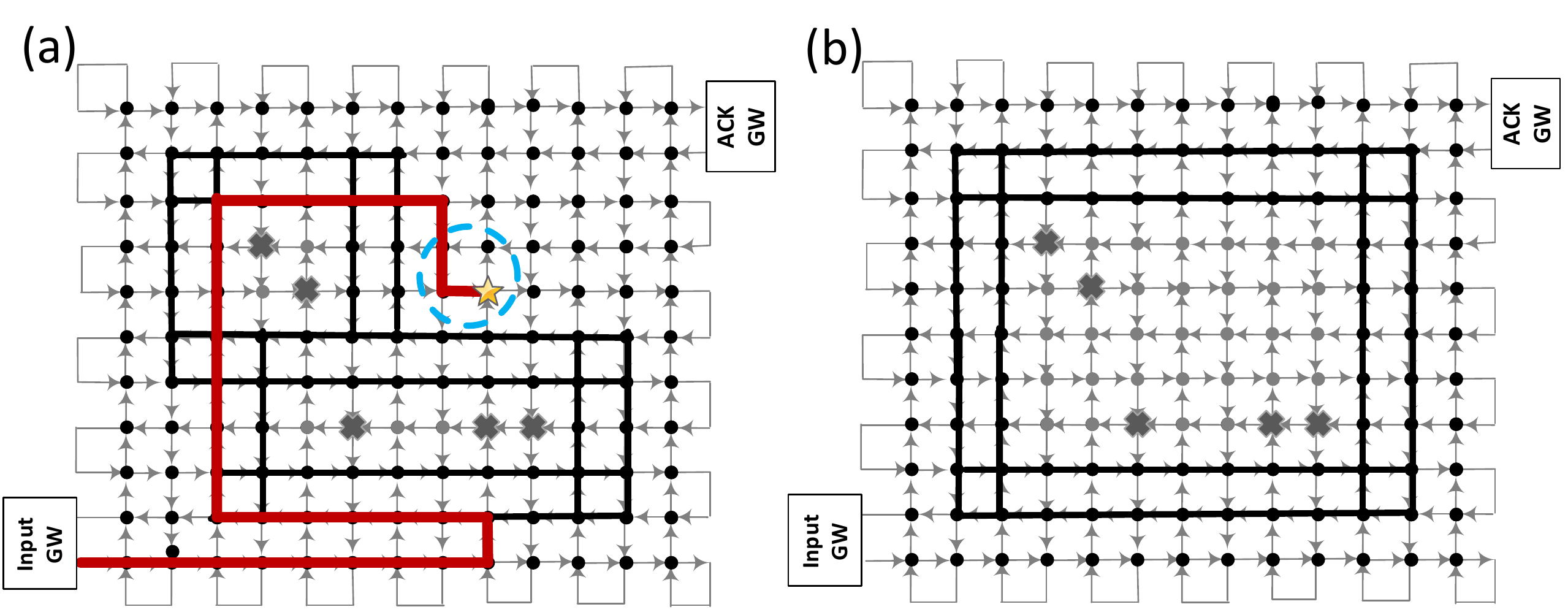}
	\caption{\small Routing turns violation: problematic overlap of faulty blocks.}
	\label{fig:over1}
\end{figure}

In some cases, the specific overlap of FBs can result in a deadlock/livelock-inducing situation. These cases occur when the northern boundary of a southern FB overlaps the southern boundary of a northern FB. In such scenario, packets break SL routing algorithm turn rules by taking a south-eastbound turn. The example of Fig.~\ref{fig:over1} shows a relevant problematic case where the routed directive makes a last south-east turn to reach its destination. In order to overcome this problem, such overlap of two FBs is eliminated by victimizing the overlapping boundary nodes, thus merging the overlapping FBs to form a so-called ``super-FB.'' This renders the combined FB a convex region, as shown in Fig. \ref{fig:over1}-(b) where said super-FB rules were applied, making SL routing now feasible. Note that any overlap between FB boundaries other than northern with southern induces no SL routing turn violations. In addition, if the overlapping FBs have the same width and are exactly on top of each other, i.e., each FB starting on the same left network column and ending on the same right network column, then such an overlap does not induce any routing deadlocks. 
\vspace{-4 mm}

\vspace{-0mm}
\section{Performance Evaluation} \label{sec:simulations}
\vspace{-1mm}



We evaluate the proposed FT routing algorithm by utilizing a custom-coded simulator built using the AnyLogic Tool platform to model the unique characteristics of the HSF CN system, i.e., its asynchronous operation. We employ a $25 \times 25$ HSF CN network (Fig. \ref{fig:TopologyRouting} shows a  $5 \times 5$ topology) consistent with the design guidelines of \cite{taghvaee2020scalability}. Beyond that, we consider all design and operational assumptions described in Section~\ref{subsec:turnRules}. Two fault distribution models are considered: Random Faults (RF) and spatially Correlated Faults (CF, or clustered) placements. RFs involve a per-node probability of failure $p_{f}$, while CFs are modeled using a spatially-dependent failure model such that nodes closer to a faulty node are more likely to also fail. To begin fault allocation under CF, a seed node is randomly chosen, while its neighboring nodes are assigned failure probabilities based on the Euclidean distance $d_{e}$ according to the function $P_{cf} = \frac{5}{d_{e}} * p_{f}$; hence a cluster of faulty nodes surrounds the seed node. Due to the math modeling differences between the CF and RF distributions, we picked distinct $0<p_{f}<1$ values for each distribution at every simulation iteration, to ensure the \textit{same percentage of total faulty nodes} in each for fair comparisons. Persistent unicast traffic is generated by the input GW, i.e., data is sequentially injected - once the first receiving node at coordinates ($x=0$,$y=0$) is freed then another packet is immediately injected into the CN. When a packet is received by a node destination, an ACK packet is generated by the \textit{preceding} node, and is then routed toward the ACK GW. In every simulation run, a packet is sent to each healthy node; $100$ such simulations are repeated for each $p_{f}$ value and fault distribution, albeit each comprising a unique faulty node pattern, and then averaged to produce a single result point. Delivery reliability is assessed in terms of the fraction of network nodes that are healthy, and as such, guaranteed by our proposed FT RA to receive packetized directives addressed to them, as they neither constitute faulty nor unsafe (i.e., victimized) nodes. Our performance metrics focus on the average routing hop count as a function of faulty node count and FB area size.
\vspace{-4 mm}
\vspace{-3mm}
\subsection{Simulation Results}\label{subsec:ExperimentalResults}
\vspace{-1mm}

\begin{figure}[t!] 
	\centering
	\vspace{-5mm}
		\includegraphics[width=.2\textwidth,trim={0 4.2cm 0 5.3cm},clip]{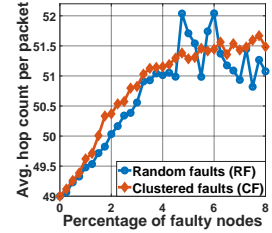}
		\vspace{-2mm}
	\caption{\small Average per-packet hop count vs. percentage of faulty nodes.}
\label{fig:hopCount}
\vspace{-5mm}
\end{figure}
\begin{figure}[t!] 
	\centering
		\includegraphics[width=.42\textwidth,trim={0 0 0 0.4cm},clip]{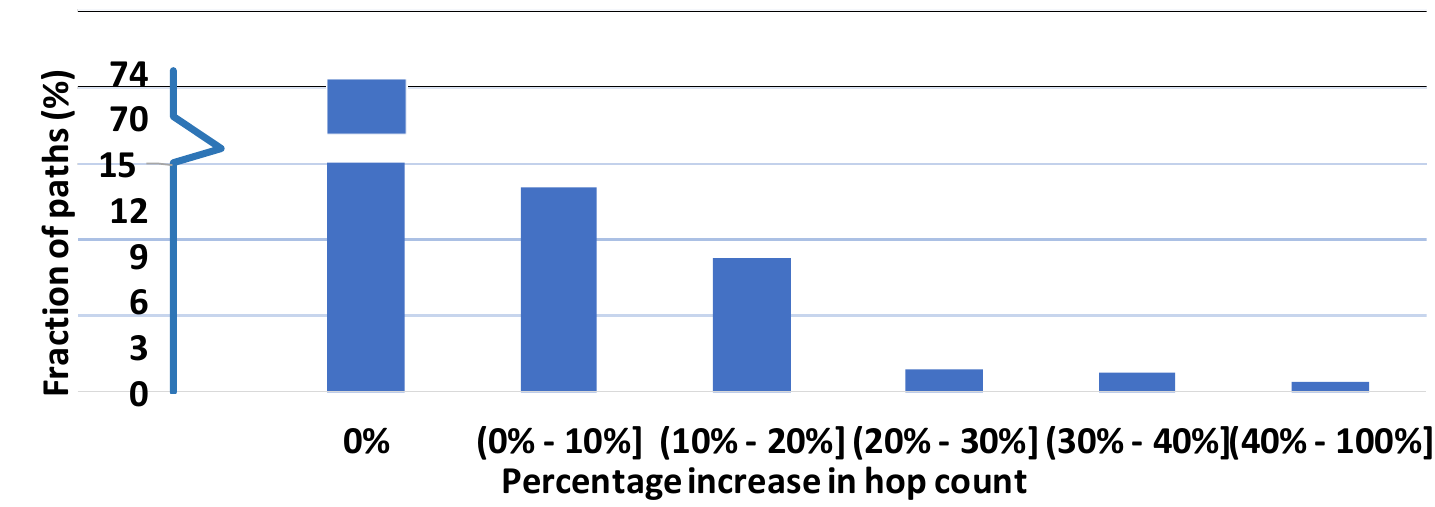}
\vspace{-2mm}
	\caption{\small Path length hop count distribution at $4\%$ random faults.}
\label{fig:hopboth}
\vspace{-5mm}
\end{figure}

Fig.~\ref{fig:hopCount} depicts the average per-packet hop count of directives \textit{delivered to healthy nodes only} increasing under the RF and CF scenarios, albeit at a rather slow pace, as a greater percentage of network nodes fail. This gradual rate is due to the fact that most route lengths remain unaffected with the presence of FBs, as they convey packets to destinations that reside north of FBs under the guidance of the proposed FT RA with a north-east biased directionality. Also, due to the location of the two GWs and said nature of routing scheme employed, only the horizontal width, and not the vertical height, of FBs contributes to the elongation of path lengths. With $4.5\%$ faulty nodes or more, the RF distribution curve wiggles due to the idiosyncrasies exhibited between FB boundary lengths, total faulty and unsafe node counts, and FB numbers and their total area size, as a result of the factorial growth of the sample space at increasing fault counts. Collectively, as Fig.~\ref{fig:hopboth} shows, FB formations impact a fraction of paths, where $\sim75\%$ of path lengths under the RF distribution with $4\%$ faulty nodes (the CF distribution behaves similarly) remain unaffected. Notably, the overall trend depicted in Fig.~\ref{fig:hopboth} was also observed at granular scales, each covering a specific percentage of faulty nodes under both the RF and CF distribution scenarios.

\begin{figure}[t!] 
	\centering
		\includegraphics[width=.4\textwidth]{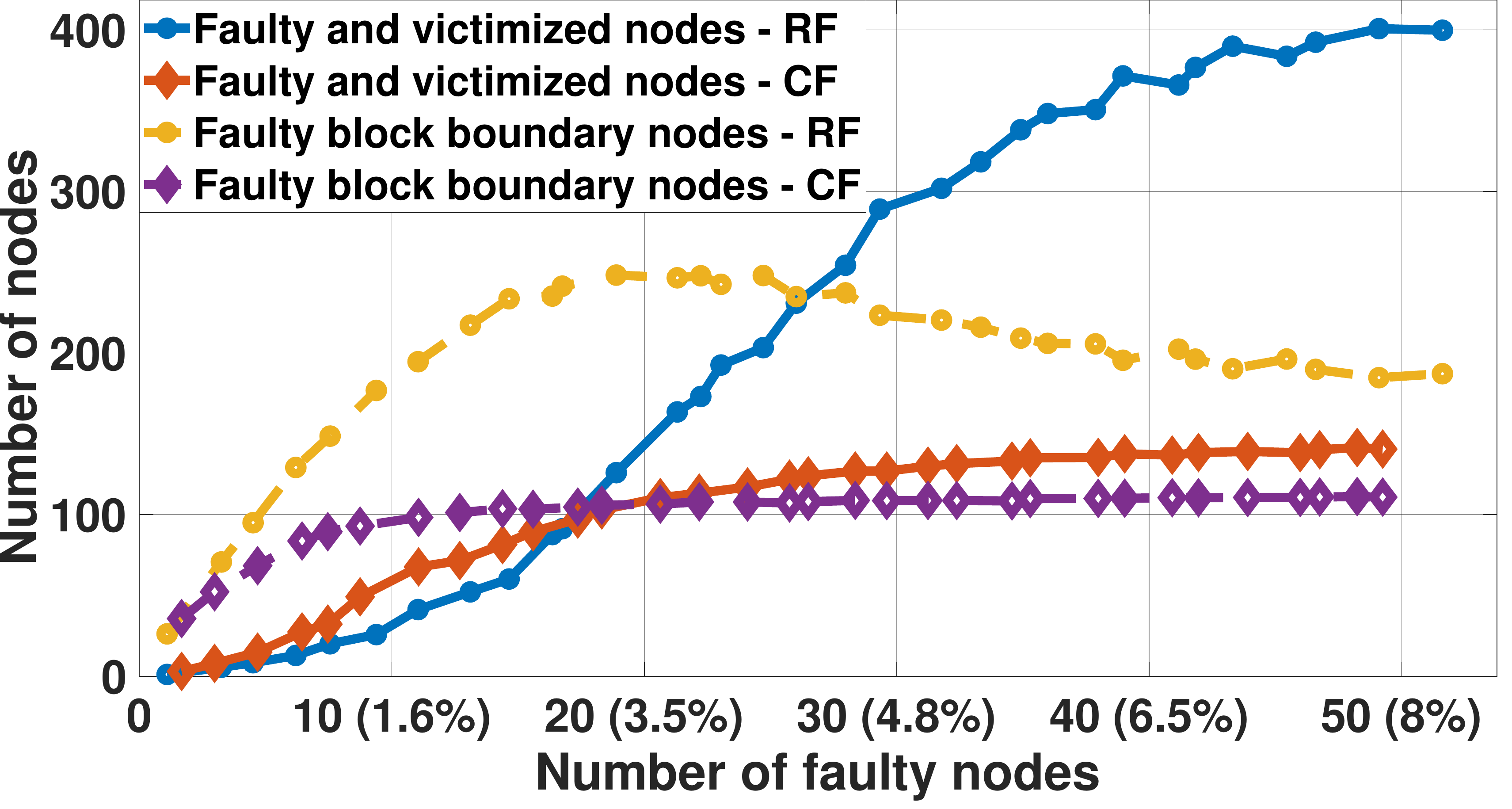}
		\vspace{-2mm}
	\caption{\small Faulty and victimized node counts, FB boundary node counts vs. total network node faults under the RF and CF scenarios.}
	\vspace{-5mm}
\label{fig:victimNumber}
\end{figure}
\begin{figure}[t!] 
	\centering
		\includegraphics[width=.42\textwidth]{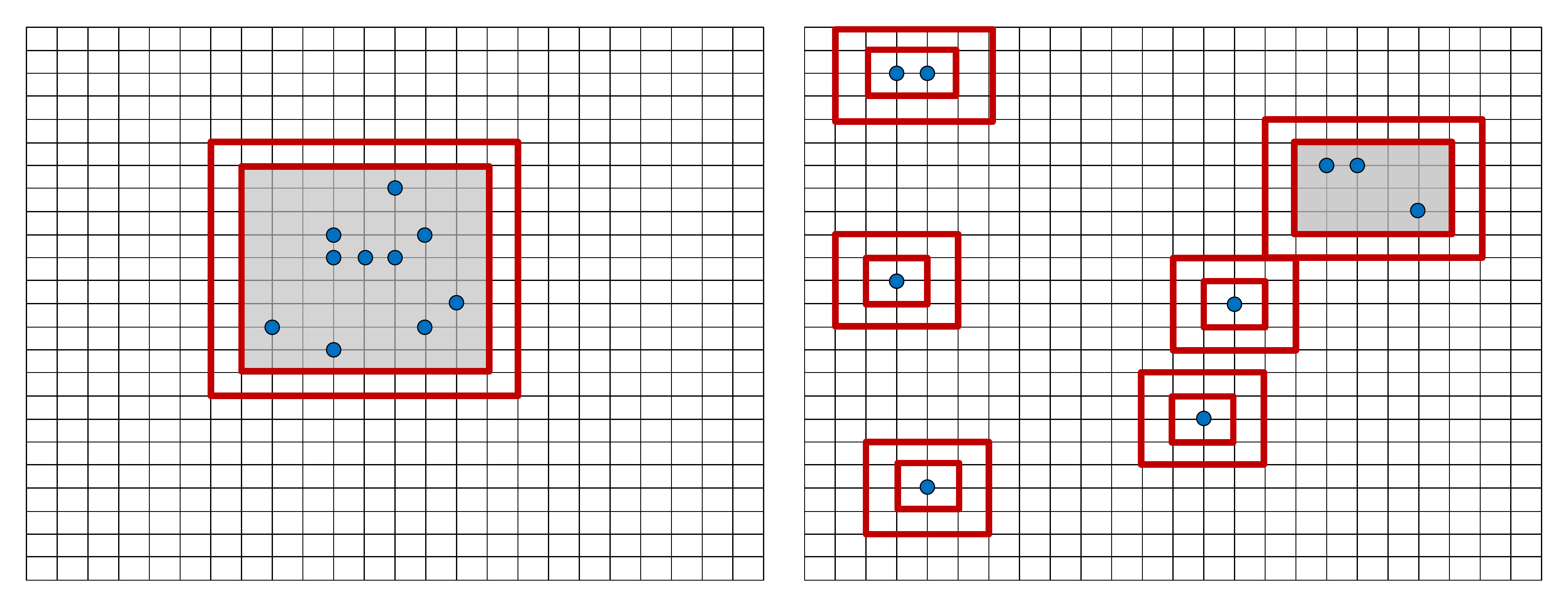}
		\vspace{-2mm}
	\caption{\small Clustered faults (left) and random faults block formations.}
	\vspace{-5mm}
\label{fig:graph_vict}
\end{figure}

The degree of victimization is assessed in Fig.~\ref{fig:victimNumber} which depicts unreachable node sums (i.e., faulty, victimized and FB boundary) as a function of faulty node counts, for the RF and CF scenarios\footnote{To maintain deadlock-freedom, the boundary nodes can convey packets but cannot serve as destination nodes (see Section~\ref{subsec:turnRules}).}. It is observed that at an average $1.6\%$ total faulty nodes, $66\%$ and $76\%$ of all nodes are spanned by the proposed FT algorithm under the respective RF and CF scenarios; these values constitute acceptable HSF functionality that is tolerant to some directives loss~\cite{taghvaee2020error}. Moreover, Fig.~\ref{fig:victimNumber} depicts the CF case performing worse in terms of the faulty and victimized node counts relative to the RF case, up until a mid-level of faulty node value of $\sim3.2\%$. This is due to the formation of a large FB(s) under CF, as opposed to more scattered and smaller FBs produced under the RF scenario which causes in-between FB manoeuvring of packets to span longer paths. This trend is further exemplified pictorially in Fig.~\ref{fig:graph_vict}, where an almost comparable number of faults for both fault models is considered: $9$ for RF and $10$ under CF. Despite this, the number of unsafe nodes differs significantly, with $46$ nodes victimized under CF, whereas only $9$ nodes become unsafe in the RFs case. However, due to the scattered pattern of FBs under RF, a larger number of boundary nodes is often produced; this backs the overall trend of the proposed FT algorithm in performing better under the CF vs. RF cases. 

\vspace{-2mm}

\section{\label{sec:conc}Conclusions and Future Work}
\vspace{-1mm}
We proposed a deterministic FT deadlock-/livelock-free routing protocol that delivers configuration directives to connected nodes in an asynchronously-operating irregular near-Manhattan HSF CN. To construct a reasonable protocol given ultra-limited resources, faults are contained in a set of disjointed rectangular regions, FBs. Simulation results demonstrate the feasibility of the proposed FT protocol under two fault distribution scenarios. Future work involves the development of enhanced routing-interconnect geometry rules to further minimize the number of victimized nodes. 
\vspace{-4 mm}

\section*{Acknowledgement}
\vspace{-2mm}
This work was supported by the European Union’s Horizon 2020 research and innovation programme-Future Emerging Topics under grant agreement No 736876 and Cyprus Research \& Innovation Foundation: HSadapt, COMPLEMENTARY/0916/0008, project.

%
\vspace{-5mm}
\bibliography{reff.bib}
\vspace{-4mm}

\end{document}